\documentclass[graybox]{svmult}

\usepackage{mathptmx}       
\usepackage{helvet}         
\usepackage{courier}        
\usepackage{type1cm}        
\usepackage{makeidx}         
\usepackage{graphicx}        
\usepackage{multicol}        
\usepackage[bottom]{footmisc}

\newcommand\lsim{\mathrel{\rlap{\lower4pt\hbox{\hskip1pt$\sim$}}
        \raise1pt\hbox{$<$}}}
        \newcommand\gsim{\mathrel{\rlap{\lower4pt\hbox{\hskip1pt$\sim$}}
        \raise1pt\hbox{$>$}}}

        \newcommand{\sfrd}{\dot{\rho}_\ast(z)}

        \newcommand{\msun}{M_\odot} \newcommand{\sneff}{\eta_{\rm SN}}
        \newcommand{\Tvir}{T_{\rm vir}} \newcommand{\zre}{z_{\rm re}}
        \newcommand{\vcirc}{v_{\rm circ}} \newcommand{\kmps}{\rm
        km~s^{-1}} \newcommand{\texp}{t_{\rm exp}}
        \newcommand{\tsurv}{t_{\rm surv}}

        \newcommand{\deltazre}{\Delta\zre}
        \newcommand{\micron}{$\mu$m}

\makeindex             

\begin{document}

\title*{Observing the First Stars and Black Holes}
\titlerunning{The First Stars and Black Holes}
\author{Zolt\'an Haiman}
\institute{Department of Astronomy, Columbia University, New York, NY 10027 {zoltan@astro.columbia.edu}}
%
%
\maketitle

\abstract{The high sensitivity of {\it JWST} will open a new window on the
end of the cosmological dark ages.  Small stellar clusters, with a
stellar mass of several $\times 10^6 {\rm M_\odot}$, and low-mass
black holes (BHs), with a mass of several $\times 10^5 {\rm M_\odot}$
should be directly detectable out to redshift $z=10$, and individual
supernovae (SNe) and gamma ray burst GRB afterglows are bright enough
to be visible beyond this redshift.  Dense primordial gas, in the
process of collapsing from large scales to form protogalaxies, may
also be possible to image through diffuse recombination line emission,
possibly even before stars or BHs are formed.  In this article, I
discuss the key physical processes that are expected to have
determined the sizes of the first star-clusters and black holes, and
the prospect of studying these objects by direct detections with {\it JWST}
and with other instruments.  The direct light emitted by the very
first stellar clusters and intermediate--mass black holes at $z>10$
will likely fall below {\it JWST}'s detection threshold.  However, {\it JWST}
could reveal a decline at the faint--end of the high--redshift
luminosity function, and thereby shed light on radiative and other
feedback effects that operate at these early epochs. {\it JWST} will also
have the sensitivity to detect individual SNe from beyond $z=10$. In a
dedicated survey lasting for several weeks, thousands of SNe could be
detected at $z>6$, with a redshift distribution extending to the
formation of the very first stars at $z\gsim 15$.  Using these SNe as
tracers may be the only method to map out the earliest stages of the
cosmic star--formation history.  Finally, we point out that studying
the earliest objects at high redshift will also offer a new window on
the primordial power spectrum, on $\sim 100$ times smaller scales than
probed by current large--scale structure data.}

\section{Introduction}
\label{sec:intro}

The formation of the first astrophysical objects and the subsequent
epoch of reionization are at the frontiers of research in Astronomy.
Recent years have seen significant progress both in our theoretical
understanding and in observational probes of this transition epoch in
the early universe.  Through a combination of methods, including
measurements of the cosmic microwave background (CMB) anisotropies,
culminating in the precise determination of the temperature and
polarization power spectra from the {\it Wilkinson Microwave
Anisotropy Probe (WMAP)} experiment
\cite{bennett03,spergel03,spergel07,page07,dunkley08,komatsu08},
Hubble diagrams of distant Supernovae \cite{riess,perlmutter}, and
various probes of large scale structure (see, e.g., references in
\cite{bahcall99,tegmark04,slosar07}) the key cosmological parameters
have been determined to high accuracy.

Because of the emergence of a concordance ($\Lambda$CDM) cosmology, we
can securely predict the collapse redshifts of the first non--linear
dark matter condensations: $2-3\sigma$ peaks of the primordial density
field on mass scales of $10^{5-6}~{\rm M_\odot}$, corresponding to the
cosmological Jeans mass, collapse at redshifts $z=15-20$.  Of course,
in addition to the cosmological parameters describing the average
background universe ($\Lambda$CDM), one also needs a description of
the seed fluctuations, to make such predictions. The nearly
scale--invariant nature of the initial fluctuation power spectrum has
also been empirically confirmed by the above--mentioned large--scale
structure data.  However, the predictions for the earliest halos rely
on extrapolating the primordial power spectrum to a wavenumber of
$k\sim 10~h{\rm Mpc^{-1}}$, a scale that is 2--3 orders of magnitude
smaller than the smallest scale directly probed by current data
(e.g. \cite{viel07} and references therein).  It is possible that the
small--scale power differs substantially from the extrapolated
value. For example, in the case of warm dark matter (WDM), the power
can be reduced by many orders of magnitude on the relevant scales,
and, to zeroth order, the first generation of halos predicted in
$\Lambda$CDM would not exist \cite{bho01}.  Recent simulations
indicate that the formation of the first halos would indeed be delayed
\cite{osheawdm,yoshidawdm}, and the details of the collapse dynamics
modified, possibly ultimately affecting the properties of the first
stars \cite{gaotheunswdm}.  Within the $\Lambda$CDM paradigm, however,
robust predictions can be made, using three--dimensional simulations
\cite{yoshida04,springel05}, rather than semi--analytical methods
\cite{ps74,smt01}. Such predictions are now limited mainly by the
$\approx 5\%$ uncertainty in the normalization of the primordial power
spectrum, $\sigma_{8h^{-1}}$ (e.g. \cite{slosar07}), which translates
essentially to a $\approx 5\%$ uncertainty in the collapse redshift of
the first structures.

On the observational side, the recent measurement of the optical depth
to electron scattering ($\tau_e=0.09 \pm 0.03$) by {\it WMAP} suggests
that the first sources of light significantly ionized the
intergalactic medium (IGM) at redshift $z\sim 11\pm 3$
\cite{page07,spergel07}.  The Sloan Digital Sky Survey (SDSS) has
uncovered a handful of bright quasars at redshifts as high as $z=6.41$
(see \cite{becker01,fan01,fan01,fan03,fan04,fan06a} or the recent
review by \cite{fan06}).  The luminosity of these sources rivals those
of the most powerful quasars at the peak of their activity at $z\simeq
2.5$.  A sizable population of $z\sim 6$ galaxies have also been
identified in the past several years (see, e.g., \cite{ellis08} for a
recent comprehensive review), many of them in broad--band photometry
in deep, wide-area {\it Hubble Space Telescope} fields
\cite{dickinson04,malhotra05,bouwens06,stanway04,stanway07}.
Narrow--band searches have also been successful in discovering
numerous $z>5$ Lyman--$\alpha$ emitting galaxies.  Starting with the
initial success of \cite{weymann98}, 10m class telescopes have now
identified about two dozen such objects out to $z\approx 6.7$ (see
\cite{mr04} for a compilation of datasets; and the more recent Subaru
survey results by, e.g., \cite{taniguchi05,kashikawa06}). In summary,
the above observational data leave little doubt that the first
galaxies and quasars were indeed well in place by redshift $z\simeq
6$, but not in significant numbers prior to redshift $z\simeq 15$.

These developments have been complemented by progress in theoretical
work focusing on the cooling and collapse of gas in primordial
low--mass halos (see reviews by \cite{zhcarn,h2review} and references
therein). A broad picture has emerged, identifying several processes
that are important in the formation of the first stars and black holes
(BHs) out of the baryonic gas in the earliest halos.  The key
ingredient in this picture is the abundance of ${\rm H_2}$ molecules
that form via gas--phase reactions in the early universe (this
beautifully simple fact was recognized as early as in 1967 by
\cite{saslaw}). The first objects form out of gas that cools via ${\rm
H_2}$ molecules, and condenses at the centers of virialized dark
matter ``minihalos'' with virial temperatures of $T_{\rm vir}\sim
200$K \cite{htl96,tetal97}. Detailed numerical simulations have shown
convergence toward a gas temperature $T\sim 300~{\rm K}\,$ and density
$n \sim 10^{4} \, {\rm cm}^{-3}$, dictated by the thermodynamic
properties of ${\rm H_2}$ \cite{ABN02,BCL02,yoshida03,yoshida08},
which allows the collapse of a single clump of mass $10^{2}-10^{3}
\,{\rm M_\odot}$ at the center of the halo. The soft UV and X--ray
radiation emitted by the stars and black holes formed in the first
handful of such clumps provide prompt and significant global feedback
on the chemical and thermal state of the IGM.  A relatively feeble
early background radiation field can already have strong effects on
gas cooling and ${\rm H_2}$ chemistry, and can produce significant
global impact, affecting the formation of the bulk of the first
astrophysical structures, and therefore the reionization history of
the IGM
\cite{hrl97,har00,mach01,mach03,oh03,rgs02a,km05,FL05,MBH06,AS07,jetal07}. The
observational data summarized above suggests that such feedback
processes indeed shaped the reionization history, and hints that
star--formation in early minihalos was suppressed \cite{hb06}.

\section{Summary of Relevant Physics at High Redshifts}
\label{sec:physics}

In the context of $\Lambda$CDM cosmologies, it is natural to identify
the first dark matter halos as the hosts of the first sources of
light.  A simplified picture for the emergence of the first sources of
light, and consequent reionization is as follows: the gas in dark
matter halos cools and turns into ionizing sources (stars and/or black
holes), which produce UV (and possibly X--ray) radiation, and drive
expanding ionized regions into the IGM.  The volume filling factor of
ionized regions then, at least initially, roughly tracks the global
formation rate of dark halos.  Eventually, the ionized regions
percolate (when the filling factor $F_{\rm HII}$ reaches unity), and
the remaining neutral hydrogen in the IGM is rapidly cleared away as
the ionizing background builds up.  However, a soft UV radiation
background at photon energies of $<13.6$eV, as well as possibly a soft
X--ray background at $\gsim 1$keV, will build up before reionization,
since the early IGM is optically thin at these energies.

We now briefly describe the various effects that should determine the
evolution of the first sources and of reionization at high redshifts.
For in--depth discussion, we refer the reader to extended reviews
(e.g. \cite{BL01}), and to reviews focusing on the roles of ${\rm
H_2}$ molecules in reionization \cite{h2review}, on the effect of
reionization on CMB anisotropies \cite{HK99}, and on progress in the
last three years in studying reionization \cite{zhcarn,ferrara07}.

In the discussions that follow, it will be useful to distinguish three
different types of dark matter halos, which can be roughly divided
into three different ranges of virial temperatures, as
follows:

\begin{tabbing}
\hspace{0.2cm} \= $300\,{\rm K}$   $\lsim$ \= $T_{\rm vir}$ \=  $\lsim  10^4\, {\rm K}$ \hspace{1cm} \= (Type II -- susceptible to ${\rm H_2}$--feedback)\\
             \> $10^4\,{\rm K}$  $\lsim$ \> $T_{\rm vir}$ \>  $\lsim 2\times 10^5\,{\rm K}$        \> (Type Ia -- susceptible to photo--heating feedback)\\
             \>                          \> $T_{\rm vir}$ \>  $\gsim 2\times 10^5\,{\rm K}$        \> (Type Ib -- gas can fall in, even in the face of photo--heating)
\end{tabbing}

We will hereafter refer to these three different types of halos as
Type II, Type Ia, and Type Ib halos.  The motivation in distinguishing
Type II and Type I halos is based on ${\rm H_2}$ molecular vs atomic H
cooling, whereas types ``a'' and ``b'' reflect the ability of halos to
allow infall and cooling of photoionized gas.  While the actual values
of the critical temperatures, and the sharpness of the transition from
one category to the next is uncertain (see discussion below), as we
argue below, star formation in each type of halo is likely governed by
different physics, and each type of halo likely plays a different role
in the reionization history.  In short, Type II halos can host
ionizing sources only in the neutral regions of the IGM, and only if
${\rm H_2}$ molecules are present; Type Ia halos can form new ionizing
sources only in the neutral IGM regions, but can do so by atomic
cooling, irrespective of the ${\rm H_2}$ abundance, and Type Ib halos
can form ionizing sources regardless of the ${\rm H_2}$ abundance, and
whether they are in the ionized or neutral phase of the IGM.  It is
useful to next provide a summary of the various important physical
effects that govern star--formation and feedback in these halos.

\begin{table*}[t]\small
\caption{\label{table:collapse} Collapse redshifts and masses of 1, 2
and 3$\sigma$ dark matter halos with virial temperatures of $T_{\rm
vir}=100$ and $10^4$K, reflecting the minimum gas temperatures
required for cooling by ${\rm H_2}$ and neutral atomic H,
respectively.}

\begin{center}
\begin{tabular}{lrrr}
$T_{\rm vir}$(K) & $\nu$ & $z_{\rm coll}$ & $M_{\rm halo}({\rm M_\odot})$ \\
\hline
100    & 1   &  7    & $2\times10^5$ \\
       & 2   &  16   & $5\times10^4$ \\
       & 3   &  26   & $3\times10^4$ \\
$10^4$ & 1   &  3.5  & $4\times10^8$ \\
       & 2   &  9    & $1\times10^8$ \\
       & 3   &  15   & $6\times10^7$ \\
\hline
\multicolumn{4}{l}{} \\
\end{tabular}\\[12pt]
\end{center}
\end{table*}

{\em ${\rm H_2}$ Molecule Formation.} In $\Lambda$CDM cosmologies,
structure formation is bottom--up: the earliest nonlinear dark matter
halos form at low masses.  Gas contracts together with the dark matter
only in dark halos above the cosmological Jeans mass, $M_{\rm
J}\approx 10^4$. However, this gas can only cool and contract to high
densities in somewhat more massive halos, with $M\gsim M_{\rm H2}
\equiv 10^5\msun[(1+z)/11]^{-3/2}$ (i.e. Type II halos), and only
provided that there is a sufficient abundance of ${\rm H_2}$
molecules, with a relative number fraction at least $n_{\rm H2}/n_{\rm
H}\sim 10^{-3}$ \cite{htl96,tetal97}.  Because the typical collapse
redshift of halos is a strong function of their size, the abundance of
${\rm H_2}$ molecules is potentially the most important parameter in
determining the onset of reionization.  For example, 2$\sigma$ halos
with virial temperatures of $100$K appear at $z=16$, while 2$\sigma$
halos with virial temperatures of $10^4$K (Type Ia halos, in which gas
cooling is enabled by atomic hydrogen lines) appear only at
$z=9$. Table~\ref{table:collapse} summarizes the collapse redshifts
and masses of dark halos at these two virial temperatures in the
concordance cosmology.  As a result, the presence or absence of ${\rm
H_2}$ in Type II halos makes a factor of $\sim$2 difference in the
redshift for the onset of structure formation, and thus potentially
effects the reionization redshift and the electron scattering opacity
$\tau$ by a similar factor. In the absence of any feedback processes,
the gas collecting in Type II halos is expected to be able to form the
requisite amount of ${\rm H_2}$ \cite{htl96,tetal97}.  However, both
internal and external feedback processes can alter the typical ${\rm
H_2}$ abundance (see discussion below).

{\em The Nature of the Light Sources in the First Halos.} A
significant uncertainty is the nature of the ionizing sources turning
on inside halos collapsing at the highest redshifts. Three dimensional
simulations using adaptive mesh refinement (AMR; \cite{ABN02}) and
smooth particle hydrodynamics (SPH; \cite{BCL02,yoshida08}) techniques
have followed the contraction of gas in Type II halos at high
redshifts to high densities. These works have shown convergence toward
a temperature/density regime of ${\rm T \sim 200~{\rm K}\,}$, ${\rm n
\sim 10^{4} \, {\rm cm}^{-3}}$, dictated by the critical density at
which the excited states of ${\rm H_2}$ reach equilibrium population
levels.  These results have suggested that the first Type II halos can
only form unusually massive stars, with masses of at least $\sim 100
\, {\rm M_\odot}$, and only from a small fraction ($\lsim 0.01$) of
the available gas.  Such massive stars are effective producers of
ionizing radiation, making reionization easier to achieve.  In
addition, it is expected that the first massive stars, forming out of
metal--free gas, have unusually hard spectra
\cite{TS2000,BKL01,schaerer02}, which is important for possibly
ionizing helium, in addition to hydrogen.  An alternative possibility
is that a similar fraction of the gas in the first Type II halos forms
massive black holes \cite{hl97,madauetal04,rog05}.  In fact, when
massive, non--rotating metal--free stars end their life, they are
expected to leave behind stellar--mass seed BHs, unless their mass is
in the range of 140-260 ${\rm M_\odot}$ \cite{Heger03}. These early
black holes can then accrete gas (possibly only after a delay -- if
the progenitor star clears the host halo of gas, the BH will have to
await until the host halo merges with another, gas--rich halo), acting
as ``miniquasars''. The miniquasars will produce a hard spectrum
extending to the soft X--rays, which could be important in catalyzing
${\rm H_2}$ formation globally (see discussion below).

{\em Efficiencies and the Transition from Metal Free to ``Normal''
Stars.} Another fundamental question of interest is the efficiency at
which the first sources inject ionizing photons into the IGM.  This
can be parameterized by the product $\epsilon_* \equiv N_\gamma f_*
f_{\rm esc}$, where $f_* \equiv M_*/(\Omega_{\rm b}M_{\rm
halo}/\Omega_m)$ is the fraction of baryons in the halo that turns
into stars; $N_\gamma$ is the mean number of ionizing photons produced
by an atom cycled through stars, averaged over the initial mass
function (IMF) of the stars; and $f_{\rm esc}$ is the fraction of
these ionizing photons that escapes into the IGM.  It is difficult to
estimate these quantities at high--redshifts from first principles,
but the discussion below can serve as a useful guide.

Although the majority of the baryonic mass in the local universe has
been turned into stars \cite{fukugita98}, the global star formation
efficiency at high redshifts was likely lower.  To explain a universal
carbon enrichment of the IGM to a level of $10^{-2}-10^{-3}~{\rm
Z_\odot}$, the required efficiency, averaged over all halos at $z\gsim
4$, is $f_*=2-20\%$ \cite{hl97}. However, the numerical simulations
mentioned in \S~2.2 above suggest that the fraction of gas turned into
massive stars in Type II halos is $f_*\lsim 1\%$.

The escape fraction of ionizing radiation in local starburst galaxies
is of order $\sim 10\%$.  The higher characteristic densities at
higher redshifts could decrease this value \cite{dsf00,wl00}, although
there are empirical indications that the escape fraction in $z\sim3$
galaxies may instead be higher \cite{spa01}, at least in some galaxies
\cite{newfesc}. Radiation can also more readily ionize the local gas,
and escape from the small Type II halos which have relatively low
total hydrogen column densities ($\lsim 10^{17}~{\rm cm^{-2}}$),
effectively with $f_{\rm esc}=1$ in the smallest halos \cite{wan04}.

The ionizing photon yield per proton for a normal Salpeter IMF is
$N_\gamma\approx 4000$. However, if the IMF consists exclusively of
massive $M\gsim 200\,{\rm M_\odot}$ metal--free stars, then $N_\gamma$
can be up to a factor of $\sim 20$ higher \cite{BKL01,schaerer02}.
The transition from metal--free to a ``normal'' stellar population is
thought to occur at a critical metallicity of $Z_{\rm cr}\sim 5\times
10^{-4}{\rm Z_\odot}$, above which cooling and fragmentation becomes
efficient and stops the IMF from being biased toward massive stars
\cite{BKL01}.  It is natural to associate this transition with that of
the assembly of halos with virial temperatures of $>10^4$K (Type Ia
halos). Type II halos are fragile, and likely blow away their gas and
``shut themselves off'' after a single episode of (metal--free)
star--formation. They are therefore unlikely to allow continued
formation of stars with metallicities above $Z_{\rm crit}$.
Subsequent star--formation will then occur only when the deeper
potential wells of Type Ia halos are assembled and cool their gas via
atomic hydrogen lines. The material that collects in these halos will
then have already gone through a Type II halo phase and contain traces
of metals.

As we argue in \S~\ref{sec:feedback} below, there exists an
alternative, equally plausible scenario.  Most Type II halos may not
have formed ${\rm any}$ stars, due to global ${\rm H_2}$
photodissociation by an early cosmic soft--UV background.  In this
case, the first generation of metal--free stars must appear in Type Ia
halos. Halos above this threshold can eject most of their
self--produced metals into the IGM, but, in difference from Type II
halos, can retain most of their gas \cite{maclowferrara99}, and can
have significant episodes of metal--free star formation.  These halos
will also start the process of reionization by driving expanding
ionization fronts into the IGM. The metals that are ejected from Type
Ia halos will reside in these photoionized regions of the IGM. As
discussed in \S~\ref{sec:feedback} below, photoionization heating in
these regions may suppress gas infall and cooling, causing a pause in
the formation of new structures, until larger dark matter halos, with
virial temperatures of $T_{\rm vir}\gsim 2\times 10^5$K (Type Ib
halos) are assembled.  The material that collects in Type Ib halos
will then have already gone through a previous phase of
metal--enrichment by Type Ia halos, and it is unlikely that Type Ib
halos can form significant numbers of metal--free stars.

\section{Feedback Effects}
\label{sec:feedback}

Several feedback effects are likely to be important for modulating the
evolution of the early cosmic star formation history and reionization.
There can be significant {\it internal} feedback in or near each
ionizing source, due to the presence of supernovae \cite{ferrara98},
or of the radiation field \cite{on99,rgs02a}, on the local ${\rm H_2}$
chemistry. The net sign of these effects is difficult compute, as it
depends on the source properties and spectra, and on the detailed
density distribution internal and near to the sources. For practical
purposes of computing the feedback that modulates the global
star--formation or reionization, we may, however, think of any
internal feedback effect as regulating the efficiency parameter
$\epsilon_*$ defined above.

Since the universe is optically thin at soft UV (below 13.6eV), and
soft X-ray ($\gsim$1keV) photon energies, radiation from the earliest
Type II halos can build up global backgrounds at these energies, and
provide prompt {\it external}, global feedback on the formation of
subsequent structures, which are easier to follow.  A very large
number of studies over the past several years have assessed the
feedback from both the LW and X--ray backgrounds quantitatively,
including many works that employed three--dimensional simulations
\cite{har00,ciardi00,mach01,mach03,rgs02a,rgs02b,glover03,km05,FL05,MBH06,AS07,jetal07}.
In isolation, Type II halos with virial temperatures as low as a few
$100$ K could form enough ${\rm H_2}$, via gas--phase chemistry, for
efficient cooling and gas contraction \cite{htl96,tetal97}. However,
${\rm H_2}$ molecules are fragile, and can be dissociated by soft UV
radiation absorbed in their Lyman-Werner (LW) bands
\cite{hrl97,ciardi00,har00,rgs01}.  In patches of the IGM
corresponding to fossil HII regions that have recombined (after the
death of short--lived ionizing source, such as a massive star), the
gas retains excess entropy for a Hubble time. This ``entropy floor''
can reduce gas densities in the cores of collapsing halos (analogously
to the case of ``preheating'' of nearby galaxy clusters), and decrease
the critical LW background flux that will photodissociate ${\rm H_2}$
\cite{oh03}.  On the other hand, positive feedback effects, such as
the presence of extra free electrons (beyond the residual electrons
from the recombination epoch) from protogalactic shocks
\cite{sk87,ferrara98}, from a previous ionization epoch
\cite{oh03,setal98}, or from X--rays \cite{hrl97,oh01,rgs02a,rgs02b},
can enhance the ${\rm H_2}$ abundance.

The extent to which star--formation in Type II halos was quenched
globally has remained unclear, with numerical simulations generally
favoring less quenching \cite{mach01,mach03,rgs02a,rgs02b,WA07} than
predicted in semi--analytical models.  Likewise, simulations find a
smaller, if any, effect from X--rays \cite{km05}. In a recent study
\cite{MBH06}, we used three-dimensional hydrodynamic simulations to
investigate the effects of a transient ultraviolet (UV) flux and a LW
background on the collapse and cooling of pregalactic clouds, with
masses in the range $10^5$ -- $10^7~{\rm M_\odot}$, at high redshifts
($z\gsim18$).  In the absence of a LW background, we found that a
critical specific intensity of $J_{\rm UV} \sim 0.1$ (in units of
$10^{-21}{\rm ergs~s^{-1}~cm^{-2}~Hz^{-1}~sr^{-1}}$) demarcates the
transition from net negative to positive feedback for the halo
population. Note that this flux is $\sim$2 orders of magnitude below
the level required for reionization (defined by the requirement of
producing a few photons per H atom). A weaker UV flux stimulates
subsequent star formation inside the fossil HII regions, by enhancing
the ${\rm H_2}$ molecule abundance. A stronger UV flux significantly
delays star--formation by reducing the gas density, and increasing the
cooling time at the centers of collapsing halos.  At a fixed $J_{\rm
UV}$, the sign of the feedback also depends strongly on the density of
the gas at the time of UV illumination.  In either case, once the UV
flux is turned off, its impact starts to diminish after $\sim30\%$ of
the Hubble time.  In the more realistic case when a permanent LW
background is present (in addition to a short--lived ionizing
neighbor), with $J_{\rm LW} \gsim 0.01 \times 10^{-21}{\rm
ergs~s^{-1}~cm^{-2}~Hz^{-1}~sr^{-1}}$, strong suppression persists
down to the lowest redshift ($z=18$) in these simulations.  The
feedback was also found to depend strongly on the mass of the Type II
halo, with the smaller halos strongly suppressed, but the larger halos
($\gsim 10^7~{\rm M_\odot}$) nearly immune to feedback.  In recent
works, \cite{WA07} and \cite{on07} found that the largest Type II
halos, with $T_{\rm vir}\gsim 4000$K, can eventually cool their gas
even in the face of a LW background with $J_{\rm LW}>0.1$, due to the
increased electron abundance and elevated temperature and cooling rate
in the inner regions of these halos.

A second type of important feedback is that photo--ionized regions are
photo-heated to a temperature of $\gsim 10^4$K, with a corresponding
increase in the Jeans mass in these regions, and possible suppression
of gas accretion onto low--mass halos
(e.g. \cite{Efstathiou92,TW96,Gnedin00b,SGB94}).  Reionization is then
expected to be accompanied by a drop in the global SFR, corresponding
to a suppression of star formation in small halos (i.e. those with
virial temperatures below $T_{\rm vir} \lsim$ $10^4$ -- $10^5$ K).
The size of such a drop is uncertain, since the ability of halos to
self--shield against the ionizing radiation is poorly constrained at
high redshifts.  Early work on this subject, in the context of dwarf
galaxies at lower redshift \cite{TW96}, suggested that an ionizing
background would completely suppress star formation in ``dwarf
galaxy'' halos with circular velocities $v_{\rm circ} \lsim~35~\kmps$,
and partially suppress star--formation in halos with 35 $\kmps$
$\lsim$ $v_{\rm circ}$ $\lsim$ 100 $\kmps$.  However, more recent
studies \cite{KI00, Dijkstra04} find that at high--redshifts ($z \gsim
3$), self-shielding and increased cooling efficiency could be strong
countering effects.  These calculations, however, assume spherical
symmetry, leaving open the possibility of strong feedback for a halo
with non--isotropic gas profile, illuminated along a low--column
density line of sight (see \cite{SIR04} for a detailed treatment of
three--dimensional gas dynamics in photo--heated low--mass halos).

Because the earliest ionizing sources formed at the locations of the
rare density peaks, their spatial distribution was strongly clustered.
Since most feedback mechanisms operate over a limited length scale,
their effects will depend strongly on the spatial distribution of
halos hosting ionizing sources. Numerical simulations are a promising
way to address feedback among clustered sources, since they capture
the full, three-dimensional relationships among the host halos
\cite{ilievsim}.  However, the dynamic range required to resolve the
small minihalos, within a large enough cosmic volume to be
representative, remains a challenge, especially in simulations that
include radiative transfer \cite{ilievRT}. Semi-analytical models
avoid the problems associated with the large dynamical range; they are
also an efficient way to explore parameter space and serve as
important sanity checks for more complicated simulations.
Semi--analytical studies to date have included {\it either} various
feedback effects
(e.g. \cite{hl98a,HH03,WL03,Cen03b,greif06,FL05,WC07,jetal07}) {\it
or} the effect of source clustering on the HII bubble--size
distribution (e.g. \cite{fzh04}), but {\it not both}.  In a recent
study \cite{roban06}, we have incorporated photo-ionization feedback,
in a simplified way, into a model that partially captures the source
clustering (i.e., only in the radial direction away from sources).
Source clustering was found to increase the mean HII bubble size by a
factor of several, and to dramatically increase the fraction of
minihalos that are suppressed, by a factor of up to $\sim 60$ relative
to a randomly distributed population. We argue that source clustering
is likely to similarly boost the importance of a variety of other
feedback mechanisms.  (This enhanced suppression can also help reduce
the electron scattering optical depth $\tau_e$, as required by the
three--year data from WMAP \cite{hb06}.)

\section{How can we detect this feedback?}

There are several ways, at least in principle, to discover the
presence of global feedback mechanisms that modulate the early
star--formation rate and reionization.  Here I simply list several
possibilities, in order to give an (admittedly crude) overview. In
\S~\ref{sec:SN}, I will discuss in more detail the one most likely to
be relevant to {\it JWST} -- tracing the cosmic star--formation history with
SNe, and searching for a feature in this ultra--high redshift version
of the ``Lilly--Madau'' diagram, caused by the feedback.

In general, the feedback processes discussed above can produce an
extended and complex (possibly even non--monotonic) reionization
history, especially at the earliest epochs.  The earliest stages of
the global reionization history (ionized fraction versus cosmic time)
can be probed in 21cm studies (see \cite{fob06} and the contributions
by Steve Furlanetto and Avi Loeb in these proceedings).  Additionally,
the measurement of polarization anisotropies by {\it Planck} can go
beyond a constraint on the total electron scattering optical depth
$\tau_e$ that is measured by {\it WMAP}, and at least distinguish
reionization histories that differ significantly
\cite{kaplinghat,HH03,wayneplanck} from each other.  Reionization may
modify the small--angle CMB anisotropies, as well, through the kinetic
SZ effect, at a level that may be detectable in the future
\cite{santos03,salvaterraksz,mcquinnksz,ilievksz,santos07}.

A recent study \cite{Oxygen1} considered the pumping of the 63.2$\mu$m
fine-structure line of neutral oxygen in the high-redshift
intergalactic medium (IGM), in analogy with the Wouthuysen-Field
effect for the 21 cm line of cosmic HI.  This showed that the soft UV
background at ~1300\AA\ can affect the fine--structure population
levels in the ground state of OI. If a significant fraction of the IGM
volume is filled with ``fossil H II regions'' that have recombined and
contain neutral OI, then this can produce a non--negligible spectral
distortion in the cosmic microwave background (CMB). A measurement of
this signature can trace the global metallicity at the end of the dark
ages, prior to the completion of cosmic reionization, and is
complementary to the cosmological 21cm studies.  In addition to the
mean spectrum, fluctuations in the OI pumping signal may be
detectable, provided the background fluctuations on arcminute scales,
around $650$GHz, can be measured to the nJy level \cite{Oxygen2}.  If
the IGM is polluted with metals at high--redshift, then CMB angular
fluctuations due to density fluctuations alone (without pumping, just
due to the geometrical effect of scattering) could also be detectable
for several metal and molecular species \cite{BHS04}.

At lower redshift, the details of the later stages of reionization can
be probed in more detail, through studying the statistics of Lyman
line absorption in the spectra of quasars (e.g. \cite{MH04,MH07} and
references therein), gamma--ray burst afterglows, and galaxies (see,
e.g., \cite{fanreview} and references therein).  In particular, the
drop in the cosmic star--formation history near the end--stages of
reionization at $z\sim 6-7$ due to photo--heating could be detected in
the high--redshift extension of the 'Lilly -- Madau' diagram
\cite{Lilly96, Madau96}, by directly counting faint
galaxies~\cite{BL00b}.  In practice, the low--mass galaxies
susceptible to the reionization suppression are faint and may fall
below {\it JWST}'s detection limit.  Whether or not these galaxies
will be detected (and in sufficient numbers so that they can reveal
the effect of reionization), depends crucially on the feedback effects
discussed above, the redshift of reionization, the size of the
affected galaxies and their typical star--formation efficiencies, as
well as the amount of dust obscuration.  Alternatively, by analyzing
the Lyman $\alpha$ absorption spectra of SDSS quasars at $z\sim 6$,
\cite{CM02} suggested, from the non-monotonic evolution of the mean
IGM opacity, that we may already have detected a drop in the SFR at
$z\sim6$.  In order to improve on this current, low signal--to--noise
result, deep, high-resolution spectra of bright quasars would be
required from beyond the epoch of reionization at $z\gsim6$ (this
could be possible with {\it JWST}, see \cite{HL99}).  A suppression of
low--mass galaxies would also increase the effective clustering of the
reionizing sources (since higher--mass halos are more strongly
clustered), and increase the fluctuations in the Lyman $\alpha$ forest
opacity at somewhat lower redshifts. This effect may already have been
observed \cite{WL06}, although the modeling details still matter, and
the presently measured scatter in opacity could perhaps still be
consistent with density fluctuations alone \cite{lof06}.

In \S~\ref{sec:SN} below, I will return to the issue of feedback, and
discuss tracing the cosmic star--formation history with distant SNe.

\section{Direct Detections of the First Sources}

The most basic question to ask is whether {\it JWST} could directly
detect the first stars or black holes.  As discussed above, the very
first star may have formed in isolation in a $10^6$ M$_\odot$ dark
matter halo at $z\gsim 15$, and it will then be beyond the reach of
direct imaging even by {\it JWST}.  The critical mass for detection
with {\it JWST} depends on the IMF and star--formation efficiency. At
$z=10$, assuming $\sim 10\%$ of the gas in a halo turns into stars,
with a normal Salpeter IMF, a 1nJy broad--band threshold at near--IR
wavelengths would allow the detection of a stellar cluster whose mass
is a several $\times 10^6$ M$_\odot$ \cite{hl97}.  It will help if the
IMF is biased toward more massive, and more luminous stars.  If the
stars were all metal--free, on the other hand, they would have a lower
flux than metal--enriched stars at {\it JWST}'s wavelengths, due to
their high effective temperatures which shifts their flux to
(observed) UV wavelengths \cite{TS2000,BKL01,schaerer02}.  The
conclusion is that while {\it JWST} is very unlikely to detect the
first individual stars directly, it will most likely directly measure
the luminosity function of faint galaxies, extending down to
sufficiently small sizes, corresponding to the halo masses
$M=10^{8-10}~{\rm M_\odot}$. This will directly probe the feedback
effects discussed above.  For example, a clear turn--over in the LF at
the luminosity corresponding to the atomic cooling threshold (or lack
of it) would be compelling evidence for ${\rm H_2}$--feedback (or lack
of it).

Likewise, one can ask whether {\it JWST} can see the first BHs
directly ?  The relevant threshold -- i.e. the lowest BH mass -- will
again depend on the spectrum and luminosity (in terms of, say, the
Eddington value) of the BHs.  With the average spectrum of quasars at
lower redshift \cite{Elvis}, and assuming Eddington luminosity, at the
1nJy threshold, {\it JWST} could detect a BH whose mass is a several
$\times 10^5$ M$_\odot$ \cite{hl98a}.  If the first BHs are the
remnants of massive, metal--free stars, then their initial masses will
be below this threshold.  Furthermore, it is not clear when such
stellar--seed BHs would start shining at a significant fraction of the
Eddington limit (since the progenitor star may clear their host halo
of gas, as mentioned above).  On the other hand, if ${\rm
H_2}$--suppression prevents most Type II halos from forming stars,
then the first stars and BHs would appear in large numbers only inside
more massive dark halos, with virial temperatures exceeding $T_{\rm
vir}\gsim 10^4$K. How gas cools, condenses, and fragments in such
halos is presently not well understood.  Using semi--analytical toy
models, \cite{oh02} argued that the initial cooling by atomic H allows
the gas to begin to collapse -- even in the face of a significant LW
background.  If the gas remained at $\sim 10^4$K, the high Jeans mass
$\sim 10^6~{\rm M_\odot}$ in these halos would suggest that a
supermassive black hole (SMBH) of a similar mass may form at the
nucleus \cite{oh02,BL03,vr05,bvr06,lodato06}. Such BHs could be
directly detectable by {\it JWST} at $z\sim 10$, provided they shine
at $\gsim 10\%$ of their Eddington luminosity.

Predictions for the number counts of high redshift galaxies and
quasars at {\it JWST}'s thresholds at near-infrared wavelengths have
been made using simple semi-analytic models \cite{hl97,hl98a}.
Surface densities as high as several sources per square arcminute are
predicted from $z~\gsim 5$, with most of these sources at $z~\gsim
10$.  These predictions, obtained from the simplest models, represent
the DM halo abundance, multiplied with optimistic star/BH formation
efficiencies, and corrected for stellar/quasar duty cycles. As such,
they could be considered as upper bounds -- feedback effects will
certainly reduce the counts. If the galaxies occupy a fair fraction
($\sim 5\%$) of the virial radius of their host halos, then a large
fraction ($\gsim 50\%$) of them can potentially be resolved with {\em
{\it JWST}}'s planned angular resolution of $\sim 0.06''$
\cite{hl98b,BL00a}.

In addition to broad--band searches, one may look for the earliest
light--sources in emission lines. The strongest recombination lines of
H and He from $5<z<20$ will fall in the near-infrared bands of {\em
{\it JWST}} and could be bright enough to be detectable.  Specific
predictions have been made for the source counts in the H$\alpha$
emission line \cite{oh01} and for the three strongest HeII lines that
could be powered either by BHs or metal--free stars with a hard
spectrum \cite{ohr01,tgs01}.  The key assumption is that most of the
ionizing radiation produced by the miniquasars is processed into such
recombination lines (rather than escaping into the IGM).  Under this
optimistic assumption (which is likely violated at least in the
smallest minihalos \cite{wan04}), the lines are detectable for a
fiducial $10^5~\msun$ miniquasar at $z=10$, or for a ``microgalaxy''
with a star--formation rate of $1 {\rm M_\odot yr^{-1}}$.  The
simultaneous detection of H and He lines would be especially
significant.  As already argued above, the hardness of the ionizing
continuum from the first sources of ultraviolet radiation plays a
crucial role in the reionization of the IGM. It would therefore be
very interesting to directly measure the ionizing continuum of any
$z>6$ source.  While this may be feasible at X-ray energies for
exceptionally bright sources, the absorption by neutral gas within the
source and in the intervening IGM will render the ionizing continuum
of high redshift sources inaccessible to direct observation out to
$\sim 1$keV.  The comparison of H$\alpha$ and HeII line strengths can
be used to infer the ratio of HeII to HI ionizing photons,
$Q=\dot{N}_{\rm ion}^{\rm HeII}/\dot{N}_{\rm ion}^{\rm HI}$.  A
measurement of this ratio would shed light on the nature of the first
luminous sources, and, in particular, it could reveal if the source
has a soft (stellar) or hard (AGN-like) spectrum.  Note that this
technique has already been successfully applied to constrain the
spectra of sources in several nearby extragalactic HII regions
\cite{getal91}.  An alternative method to probe the hardness of the
spectrum of the first sources would be to study the thickness of their
ionization fronts \cite{SIR04,ZS2005,roban07}; this may be possible
for bright quasars in Lyman line absorption \cite{roban07} or in 21cm
studies \cite{ZS2005,TZ2007,roban07}.

Provided the gas in the high redshift halos is enriched to near--solar
levels, several molecular lines may be visible. In fact, CO has
already been detected in the most distant $z=6.41$ quasar
\cite{wbetal03}, and, in fact, spatially resolved \cite{wbetal04}.
The detectability of CO for high redshift sources in general has been
considered by \cite{ss97} and by \cite{gss01}.  For a star formation
rate of $\gsim 30~\msun/$~yr, the CO lines are detectable at all
redshifts $z=5-30$ by the Millimeter Array (the redshift independent
sensitivity is due to the increasing CMB temperature with redshift),
while the Atacama Large Millimeter Array (ALMA) could reveal fainter
CO emission.  The detection of these molecular lines will provide
valuable information on the stellar content and gas kinematics in the
earliest halos.

Finally, the baryons inside high--redshift halos with virial
temperatures $T\gsim10^4$K need to cool radiatively, in order to
condense inside the dark matter potential wells, even before any stars
or black holes can form.  The release of the gravitational binding
energy, over the halo assembly time--scale, can result in a
significant and detectable Ly$\alpha$ flux \cite{hsq00,Fardal01}.  At
the limiting line flux $\approx 10^{-19}~{\rm
erg~s^{-1}~cm^{-2}~asec^{-2}}$ of {\it JWST}, several sufficiently
massive halos, with velocity dispersions $\sigma\gsim 120~{\rm
km~s^{-1}}$, would be visible per $4^\prime\times4^\prime$ field.  The
halos would have characteristic angular sizes of $\approx 10$\H{},
would be detectable in a broad--band survey out to $z\approx 6-8$, and
would provide a direct probe of galaxies in the process of
forming. They may be accompanied by He$^+$ Ly$\alpha$ emission at the
$\approx 10\%$ level \cite{hecool}, but remain undetectable at other
wavelengths. The main challenge, if such blobs are detected, without
any continuum source, will likely be the interpretation -- as is the
case for the currently detected extended Lyman $\alpha$ sources (there
are currently $\sim $three dozen extended Ly$\alpha$ blobs known, with
1/3rd of such objects in the largest sample \cite{Matsuda04}
consistent with being powered by cooling radiation).

Monte Carlo calculations of Ly$\alpha$ radiative transfer through
optically thick, spherically symmetric, collapsing gas clouds were
presented in \cite{DijkstraLya1,DijkstraLya2}, with the aim of
identifying a clear diagnostic of gas infall (a similar effort in 3D
is being carried out \cite{Tasitsiomi}). These represent simplified
models of proto--galaxies that are caught in the process of their
assembly.  Such galaxies can produce Ly$\alpha$ flux over an extended
solid angle, either from a spatially extended Ly$\alpha$ emissivity,
or from scattering effects, or both.  We presented a detailed study of
the effect of the gas distribution and kinematics, and of the
Ly$\alpha$ emissivity profile, on the emergent spectrum and surface
brightness distribution. The emergent Ly$\alpha$ spectrum is typically
double--peaked and asymmetric. The surface brightness distribution is
typically flat, and the detection of a strong wavelength dependence of
its slope (with preferential flattening at the red side of the line)
would be a robust indication that Ly$\alpha$ photons are being
generated (rather than just scattered) in a spatially extended region
around the galaxy, as in the case of a cooling flow. An alternative,
robust diagnostic for scattering is provided by the polarization of an
extended Lyman $\alpha$ source \cite{Dijkstrapol}. Spectral
polarimetry (in particular, the wavelength-dependence of the
polarization) can differentiate between Lya scattering off infalling
gas and outflowing gas.

\section{Tracing The Beginning of the Cosmic Star--Formation History With Supernovae}
\label{sec:SN}

As argued above, the very first stellar clusters may fall below the
direct detection of {\it JWST}.  A promising way (and probably the
only way) to directly observe the first stars individually is through
their explosions, either as Supernovae or gamma ray bursts (GRBs).
Before we present the expectations for SNe in detail, we first briefly
discuss an alternative, GRB afterglows.  In the years leading up to
the launch of the {\it Swift} satellite,\footnote{See
http://swift.gsfc.nasa.gov} it has been increasingly recognized that
distant gamma ray bursts (GRBs) offer a unique probe of the high
redshift universe.  In particular, GRBs are the brightest known
electromagnetic phenomena in the universe, and can be detected up to
very high redshifts (e.g. \cite{wijers98,lr00,cl00}), well beyond the
redshift $z\approx 6.5$ of the most distant currently known quasars
\cite{fan06} and galaxies \cite{ellis08}.  There is increasing
evidence that GRBs are associated with the collapse of short--lived,
massive stars, including the association of bursts with star--forming
regions (e.g. \cite{fruchter99}, a contribution of supernova light to
the optical afterglow (e.g. \cite{bloom99,garnavich03}), and most
directly, association with a supernova \cite{stanek03,hjorth03}.

As a result, the redshift distribution of bursts should follow the
mean cosmic star--formation rate (SFR).  Several studies have computed
the evolution of the expected GRB rate under this assumption, based on
empirical models of the global SFR
\cite{totani97,totani99,wijers98,lr00,cl00}.  Determinations of the
cosmic SFR out to redshift $z\sim 5$
(e.g. \cite{Bunker04,Gabasch04,Giavalisco04}) have shown that
star--formation is already significant at the upper end of the
measured redshift range, with $\gsim 10\%$ of all stars forming prior
to $z=5$, which would result in a significant population of GRBs at
these redshifts.  Further associating star--formation with the
formation rate of non--linear dark matter halos, and using theoretical
models based on the Press--Schechter formalism \cite{ps74}, it is
possible to extrapolate the SFR and obtain the GRB rates expected at
still higher redshifts \cite{BL02,CS02,MPH05}. These studies have
concluded that a significant fraction (exceeding several percent) of
GRBs detected at {\it Swift}'s sensitivity should originate at
redshifts as high as $z>10$.  The spectra of bright optical/IR
afterglows of such distant GRBs can provide information both on the
progenitor, and also reveal absorption features by neutral hydrogen in
the intergalactic medium (IGM), and can serve as an especially clean
probe of the reionization history of the universe
\cite{jordi98,lr00,CS02,lh03,BL04_grbvsqso}.

In summary, the main advantage of GRBs is that they are bright and can
be seen essentially from arbitrary far away. The main drawback,
however, is that they are rare.  Even the most optimistic among the
above--listed models predict only a handful of detectable GRBs per
year from $z>6$ -- indeed, in the past two years, only one GRB has
been discovered at $z>6$ (GRB050904 at $z=6.3$; \cite{kawai06}).

\subsection{Supernovae From The First Stars}

Although SNe are not as bright as GRB afterglows, they are bright
compared to galaxies at the very faint end of the luminosity function,
and individual core--collapse SNe would still be visible from beyond
$z=10$.  Furthermore, since they occur much more frequently than GRBs,
and they remain visible for a longer time than GRB afterglows, they
will likely offer a better (and possibly only) chance to trace out the
cosmic star formation history accurately, with a statistically
significant sample extending beyond $z>10$.  The above is true even if
the bulk of early supernovae are similar to the core--collapse SNe in
the local universe (note that Type Ia supernovae will not have time to
form at $z \gg 6$). In addition, the pair--instability supernovae from
massive, metal--free stars are expected to be much brighter than Type
II supernovae from normal (metal--enriched) stars \cite{HW02}.

There have been several studies of the expected early supernova rate
(SNR), calibrated to the observed metallicity of the Lyman $\alpha$
forest \cite{MeR97} or the observed SFR at lower redshifts
\cite{DF99}. The expected rate of pair--instability SNe from very
high-$z$ from the first generation of metal--free stars was studied by
\cite{WA05}.  In a recent work \cite{MJH06}, we constructed the
expected detection rate of high-$z$ SNe in SNe surveys for {\it {\it
JWST}}. We also quantified the prospects of detecting a drop in the SN
rate due to photo--heating feedback at reionization.  Given that SNe
may be our only hope to directly map out the beginning of the cosmic
star--formation history, most of the rest of this article is devoted
to discussing this possibility in detail.

\subsection{The Global Star Formation and Supernova Rates}

The global SFR density can be obtained and extrapolated by a common
approach, based on the dark matter halo formation rate, calibrating
the star--formation efficiency to the SFR at redshift $z\lsim 6$.
From this extrapolated SFR density, we can then obtain the intrinsic
supernova rate by using the properties of core--collapse SNe in the
local universe as a guide.

In particular, we can estimate the global SFR density at redshift $z$ as
\begin{equation}
\label{eq:sfrd_general}
\sfrd = \epsilon_\ast \frac{\Omega_b}{\Omega_{\rm M}} \int_{M_{\rm min}(z)}^{\infty} dM \int_{\infty}^{z} dz' M \frac{\partial^2 n(>M, z')}{\partial M \partial z'} P(\tau) ~ ,
\end{equation}
where $\epsilon_\ast$ is the efficiency (by mass) for the conversion
of gas into stars, $M dM (\partial n(>M, z) / \partial M)$ is the mass
density contributed by halos with total (dark matter + baryonic)
masses between $M$ and $M + dM$, $t(z)$ is the age of the universe at
redshift $z$, and $P(\tau)$ is the probability per unit time that new
stars form in a mass element of age $\tau \equiv t(z) - t(z')$
(normalized to $\int_0^\infty d\tau P(\tau)=1$). We adopt the fiducial
value of $\epsilon_\ast = 0.1$ (see, e.g., \cite{Cen03b}). Note that
the star formation efficiency in minihalos (i.e. halos with virial
temperatures below $10^4$K) that contain pristine metal--free gas
could be significantly lower than 0.1, as suggested by numerical
simulations of the first generation of stars \cite{ABN02, BCL02}. The
expected pre--reionization SNe rates would then lie closer to our
$T_{\rm vir} \gsim 10^4$ K curves prior to reionization (see
discussion below), making the detection of the reionization feature
significantly more difficult.  However, the efficiency is likely to be
very sensitive to even trace amounts of metallicity~\cite{BFCL01}, and
conditions for star--formation may result in a standard initial mass
function (IMF) in gas that has been enriched to metallicities above a
fraction $10^{-4}$ of the solar value.  Indeed, it is unlikely that
metal--free star--formation in minihalos can produce enough ionizing
photons to dominate the ionizing background at reionization
(e.g. \cite{har00,HH03,SYAHS04}).  We assume further that
star--formation occurs on an extended time-scale, corresponding to the
dynamical time, $t_{\rm dyn} \sim [ G \rho(z) ]^{-1/2}$ \cite{CO92,
Gnedin96}:
\begin{equation}
P(\tau) = \frac{\tau}{t_{\rm dyn}^2} \exp \left[ -\frac{\tau}{t_{\rm dyn}} \right] ~ ,
\end{equation}
where $\rho(z) \approx \Delta_c \rho_{\rm crit}(z)$ is the mean mass
density interior to collapsed spherical halos (e.g. \cite{BL01}),
and $\Delta_c$ is obtained from the fitting formula in \cite{BN98},
with $\Delta_c$ = $18\pi^2$ $\approx$ 178 in the Einstein--de Sitter
model.  The minimum mass, $M_{\rm min}(z)$ in eq. (1), depends on the efficiency
with which gas can cool and collapse into a dark matter halo.  Prior
to reionization and without molecular hydrogen, $M_{\rm min}(z)$
corresponds to a halo with virial temperature, $T_{\rm vir} \sim 10^4$
K; with a significant $\rm H_2$ abundance, the threshold decreases to
$T_{\rm vir} \sim 300$ K (\cite{har00}; we use the conversion
between halo mass and virial temperature as given in \cite{BL01}).
Post reionization, the Jeans mass is raised, so $M_{\rm min}(z)$ could
increase.  The degree of self-shielding, the ability of the halo gas
to cool, as well as the amount of $\rm H_2$ present in the
high--redshift low--mass halos is uncertain~\cite{Dijkstra04}, and so
below we present results for several values of $M_{\rm min}(z)$, which
we will henceforth express in terms of $T_{\rm vir}$(z).

Next, from $\sfrd$ we obtain the intrinsic differential SNR (number of
core collapse SNe per unit redshift per year) with
\begin{equation}
\label{eq:SNint}
\frac{d\dot{N}}{dz} = \sneff \frac{1}{1+z} \frac{dV(z)}{dz} \sfrd ~ ,
\end{equation}
where the factor $1/(1+z)$ accounts for time dilation, $dV(z)/dz$ is
the comoving volume in our past light cone per unit redshift, and
$\sneff$ is the number of SNe per solar mass in stars.  For a fiducial
Salpeter initial mass function (IMF), we obtain $\sneff \sim
1/180~{\rm M_\odot^{-1}}$, assuming all stars with masses $9~\msun
\lsim M \lsim 40~\msun$ become core collapse SNe \cite{Heger03}. We
neglect the lifetime of these high mass stars in determining our SNR;
this is a reasonable assumption as the lifetimes (as well as the
spread in the lifetimes) are shorter than a unit redshift interval for
redshifts of interest.  Note that an alternative extreme shape for the
IMF, consisting entirely of 100--200 $\msun$ stars \cite{ABN02,
BCL02} would yield a similar value for $\sneff$.

\begin{figure}[t]
\centerline{\includegraphics[scale=.45]{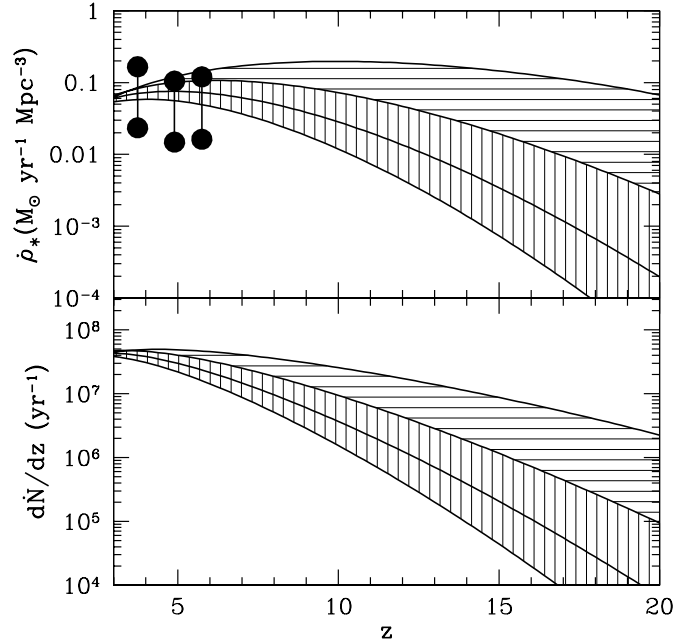}}
\caption{{\it Upper Panel}: SFR densities obtained in a model based on
dark matter halo abundances.  The curves ({\it top to bottom})
correspond to different lower cutoffs on the virial temperatures of
star--forming halos, $\Tvir \gsim 300,~10^4,~4.5\times 10^4$, and $1.1
\times 10^5$ K (corresponding to circular velocity thresholds of
$\vcirc \gsim 3,~17,~35$, and $55~\kmps$).  Dots indicate results from
GOODS \cite{Giavalisco04}: the lower set of points assume no dust
correction, while the upper set of points are dust corrected;
statistical 1-$\sigma$ error bars lie within the points.  The lines
connecting each pair of points span the expected range of SFR
densities.  {\it Lower Panel}: the total global supernova rate
accompanying the SFR densities in the top panel (adapted from
\cite{MJH06}).
\label{fig:SFR_SNint}}
\end{figure}

We present our SFR densities ({\it top panel}) and SNRs ({\it bottom
panel}) in Figure~\ref{fig:SFR_SNint}.  The curves correspond to
redshift--independent virial temperature cutoffs of $\Tvir =
300,~10^4,~4.5\times 10^4$, and $1.1 \times 10^5$ K (or circular
velocities of $\vcirc = 3,~17,~35$, and $55~\kmps$, respectively), top
to bottom, spanning the expected range~\cite{TW96, Dijkstra04}.  Also
shown are results from GOODS \cite{Giavalisco04}: the bottom
points assume no dust correction and the top points are dust corrected
according to \cite{AS00}; the statistical error bars lie within the
points \cite{AS00}.  As there are large uncertainties associated with
dust correction, each pair of points (top and bottom) serves to
encompass the expected SFR densities.

Our SFRs are consistent with other theoretical predictions
(e.g. \cite{SPF01,BL00b, BL02}), as well as other estimates
from the Hubble Ultra Deep Field \cite{Bunker04} and the FORS Deep
Field on the VLT \cite{Gabasch04}, after they incorporate a factor of
5--10 increase in the SFR \cite{AS00} due to dust obscuration.
Furthermore, we note that our SNRs, which at $z\gsim5$ yield 0.3 -- 2
SNe per square arcminute per year, are in good agreement with the
$\sim$ 1 SN per square arcminute per year estimated by \cite{MeR97}
by requiring that high-redshift SNe produce a mean metallicity of $\sim
0.01~Z_\odot$ by $z \sim 5$.  Note, however, that the rates we obtain
are significantly higher (by a factor of $\sim$ 60 -- 2000 at
$z\sim20$) than those recently found by~\cite{WA05}.  The reason for
this large difference is that Wise \& Abel consider star--formation
only in minihalos, and they assume a very low star--formation
efficiency of a single star per minihalo, as may be appropriate for
star--formation out of pristine (metal--free) gas in the first
generation of minihalos~\cite{ABN02, BCL02}. In contrast, we assume
an efficiency of $\epsilon_\ast=0.1$, which may be more appropriate
for star--formation in pre--enriched gas that dominates the SFR just
prior to reionization (including star--formation in minihalos).

As mentioned above, by increasing the cosmological Jeans mass,
reionization is expected to cause a drop in the SFR (and hence the
SNR), with the rates going from the horizontally striped region in
Figure~\ref{fig:SFR_SNint} at $z > \zre$ to the vertically striped
region at $z < \zre$.

The redshift width of this transition is set by a combination of
large-scale cosmic variance, radiative transfer, and feedback effects.
For the majority of the paper, we use $\deltazre \sim 1$ as a rough
indicator of the width of the transition we are analyzing.  We
distinguish between ``reionization'' and a ``reionization feature'',
and use $\deltazre$ as an indicator of the width of the later.  Even
with an extended reionization history ($\Delta z \sim 10$), fairly
sharp ($\deltazre \lsim 3$) features are likely, as discussed in
detail in \cite{MJH06}.

The other important factor determining the usefulness of the method
proposed here is the factor by which the SFR drops during the
reionization epoch.  The size of this drop is mediated by the
effectiveness of self-shielding and gas cooling during photo--heating
feedback: i.e. on whether or not the star--formation efficiency is
significantly suppressed in those halos that dominate the SFR and SNR
immediately preceding the reionization epoch.  Given the uncertainties
about this feedback discussed above, we will consider a range of
possibilities below, parameterized by the modulation in the virial
temperature threshold for star--formation during reionization.


\subsection{The Rate of High--Redshift SNe Detectable in a Future Survey}
\label{sec:detection}

Given the intrinsic star--formation and SN rates, our next task is to
estimate the number of SNe that could actually be detected in a future
search with {\it JWST}.  In general, the number of SNe per unit redshift,
$dN_{\rm exp}/dz$, that are bright enough to be detectable in an
exposure of duration $\texp$ can be expressed as
\begin{equation}
\label{eq:SNexp}
\frac{dN_{\rm exp}}{dz} = \frac{d\dot{N}}{dz} \int_{0}^{\infty} f_{\rm SN}(>t_{\rm obs}) ~  dt_{\rm obs} ~ ,
\end{equation}
where $(d\dot{N}/dz) dt_{\rm obs}$ is the number of SNe which occurred
between $t_{\rm obs}$ and $t_{\rm obs} + dt_{\rm obs}$ ago (per unit
redshift; note that the global mean SNR will evolve only on the Hubble
expansion time--scale, and can be considered constant over several
years), and $f_{\rm SN}(>t_{\rm obs})$ is the fraction of SNe which
remains visible for at least $t_{\rm obs}$ in the observed frame.
Then the total number of SNe detected in a survey of duration $\tsurv$
is
\begin{equation}
\label{eq:SNobs}
N_{\rm surv} = \frac{\tsurv}{2~\texp} \frac{\Delta\Omega_{\rm FOV}}{4 \pi} N_{\rm exp} ~ ,
\end{equation}
where $\Delta\Omega_{\rm FOV}$ is the instrument's field of view, and
$\tsurv/(2\texp)$ is the number of fields which can be tiled in the
survey time, $\tsurv$ (we add a factor of 1/2 to allow for a second
pair of filters to aid in the photometric redshift determination; note
that this provides for imaging in 4 different {\it JWST} bands; see
discussion below).  Note also that equation (\ref{eq:SNobs}) is
somewhat idealized, in that it assumes continuous integration for a
year, and e.g., does not account for time required to slew the
instrument to observe different fields. In principle, each field has
to have repeated observations (to detect SNe by their variability),
and therefore any dedicated survey should target fields that have
already been observed. Furthermore, a dedicated year--long program may
not be necessary, because the effect could be detected with relatively
few fields (at least under optimistic assumptions; see discussion
below), and several fields with repeated imaging (separated by $>$ 1
-- 2 years) may already be available from other projects; these fields
can then be used for the SN search.

In general, $f_{\rm SN}(>t_{\rm obs})$ in equation~(\ref{eq:SNexp}), i.e. the
fraction of SNe which remain visible for at least $t_{\rm obs}$, depends on (i)
the properties of the SN, in particular their peak magnitude and
lightcurve, and the distribution of these properties among SNe, and
(ii) on the properties of the telescope, such as sensitivity, spectral
coverage, and field of view.  In the next two subsections, we discuss
our assumptions and modeling of both of these in turn.

\subsubsection{Empirical Calibration of SN Properties}
\label{sec:empirical}

At each redshift, we run Monte-Carlo simulations to determine $f_{\rm
SN}(>t_{\rm obs})$ in equation~(\ref{eq:SNexp}). We use the observed properties
of local core--collapse SNe (CCSNe) in estimating $f_{\rm SN}(>t_{\rm obs})$.
For the high redshifts of interest here, we only consider core collapse SNe of Type II.
SNe resulting from the collapse of Chandrasekhar--mass white dwarfs (Type
Ia) are expected to be extremely rare at high redshifts ($z \gsim 6$), as
the delay between the formation of the progenitor and the SN event ($\gsim
1$ Gyr; \cite{Strolger04}) is longer than the age of the universe at these
redshifts.  Local core CCSNe come in two important varieties, types
IIP and IIL, differentiated by their lightcurve shapes. We ignore the
extremely rare additional CCSN types, e.g Type IIn and IIb, which
appear to have significant interaction with circumstellar material and
constitute less than 10\% of all CCSNe. Type Ib/c, which may or may not also
result from core collapse, have luminosities and light-curves that are
similar to Type IIL and occur less frequently. While the relative numbers
of Type IIP and Type IIL SNe are not known even for nearby SNe, 
estimates imply that they are approximately equal in frequency
\cite{C97}. We therefore assume that 50\% of the high-redshift SNe
are Type IIP and 50\% are Type IIL.

CCSNe result from the collapse of the degenerate cores of high-mass
stars.  The luminosity of CCSNe is derived from the initial shock
caused by the core-collapse which ionizes material and fuses unstable
metal isotopes (see \cite{LS03} and references therein for a more
detailed description of SN lightcurves).  In the early stages of the
SN, the shock caused by the core collapse breaks out from the surface
of the progenitor (typically high mass red giants), resulting in a
bright initial peak in the light curve that lasts less than a few days
in the rest--frame of the SN.  As the shock front cools, the SN
dims. However, the SN may then reach a plateau of constant luminosity
in the light curve, believed to be caused by a wave of recombining
material (ionized in the shock) receding through the envelope.  The
duration and strength of this plateau depends on the depth and mass of
the progenitor envelope, as well as the explosion energy, with those
SNe exhibiting a strong plateau classified as Type IIP. A typical
plateau duration is $\lsim 100$ rest--frame days \cite{P94}. In Type
IIL SNe, this plateau is nearly non-existent, and the lightcurve
smoothly transitions from the rapid decline of the cooling shock to a
slower decline where the luminosity is powered by the radioactive
decay of metals in the SN nebula.  After the plateau, Type IIP SNe
also enter this slowly declining `nebular' phase.

These observationally determined behaviors have been summarized in a
useful form as lightcurve templates in \cite{DB85}. We use these
template lightcurves in determining $f_{\rm SN}(>t_{\rm obs})$, and normalize
the lightcurves using Gaussian--distributed peak magnitudes
(i.e. log--normally distributed in peak flux) determined by
\cite{Richardson02} from a large sample of local Type IIP and Type
IIL SNe (see Table~\ref{tbl:peaks}). We perform the Monte-Carlo
simulations with both the dust corrected, and dust uncorrected values
in \cite{Richardson02}, since the dust production history of the
early universe is poorly understood and is essentially unconstrained
empirically.

We use a combined high-resolution HST STIS + ground--based spectrum of
the Type IIP supernova SN1999em (the `November 5th' spectrum of
\cite{Baron00}) as the template SN spectrum in order to obtain
$K$-corrections (with the \cite{DB85} lightcurves given in the
restframe $B$ filter).  This spectrum was obtained within 10 days of
maximum light, i.e. during the initial decline of the SN brightness,
and has been dereddened by A$_{V}=0.3$ mag \cite{Baron00,H01}.  While
the spectrum, and hence the $K$-corrections, of SNe evolve during the
lightcurve, this effect is not strong for the wavelengths of interest
\cite{P94}, especially since the lightcurve template we use is well
matched to the wavelengths being probed by the observations we
consider below (leading to small $K$-corrections).  We have also used
this Type IIP SN template spectrum to calculate $K$-corrections for
Type IIL SNe.  This is necessary due to the lack of restframe UV
spectra of Type IIL SNe that can be combined with optical spectra, and
justifiable because the $K$-corrections are relatively small and the
broadband colors of both Type IIP and IIL SNe (a measure of the
spectral shape that determines the $K$-corrections) are similar, at
least in the optical \cite{P94}.  Note that we assume that very-high
redshift core-collapse SNe are similar to local SNe in their spectra,
peak luminosities, and temporal evolution. However, these assumption
do not significantly impact our conclusions below, as long as (i) the
average properties of the SNe do not change \emph{rapidly} at high
redshift (which could mimic the reionization drop), (ii) they do not
become preferentially under-luminous (which would make high-$z$ SNe
less detectable, lowering the statistical confidence at which the
reionization drop is measured), and (iii) the detection efficiency
does not change rapidly with redshift (e.g. due to instrument
parameters or spectral lines).


\begin{table}[ht]
\caption{Means and standard deviations of the adopted peak absolute
magnitudes of core collapse SNe.  Values are taken from
\cite{Richardson02}. Note: an $M_B=-17$ SN would be detectable out to
$z \approx 8.2$ at the flux threshold of 3 $\rm nJy$ in the 4.5 $\mu$$\rm
m$ band with {\it JWST}. }
\vspace{-0.8cm}
\label{tbl:peaks}
\begin{center}
\begin{tabular}{ccccc}
\\
\hline
\hline
  & \multicolumn{2}{c}{Corrected for Dust} & \multicolumn{2}{c}{Not Corrected for Dust}\\
\hline
SN Type & $\langle M_{B} \rangle$ & $\sigma$ & $\langle M_{B} \rangle$ & $\sigma$\\
\hline
\hline
IIP & -17.00 & 1.12 & -16.61 & 1.23\\
IIL & -18.03 & 0.9 & -17.80 & 0.88\\
\hline
\hline
\end{tabular}\\
\end{center}
\end{table}


In order to use the method outlined above to probe reionization, the
SN redshifts must also be known to an accuracy of $\Delta z\lsim 1$.
SN redshifts can be determined via spectroscopy of either the SN
itself, or of the host galaxy.  However, as we have already noted, the
host galaxies may only be marginally detectable even in imaging, and
the SNe may be too faint for anything other than extremely low
resolution spectroscopy.  We present here only a very brief example of
the possibility of obtaining redshifts from the extremely low
resolution ($ \lambda /\Delta \lambda \sim 5$) spectra provided by
multi-band imaging: a complete investigation of this possibility is
warranted, but is beyond the scope of this paper.

\begin{figure}[t]
\centerline{\includegraphics[scale=.45]{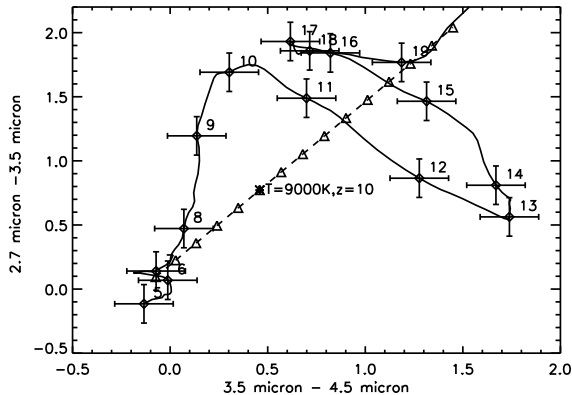}}
\caption{Infrared colors of a Type II SN as a function of its
redshift.  \emph{Solid curve}: The template spectrum of the text, from
the Type IIP SN 1999em. Diamonds are placed at intervals of $\Delta
z=1$, labeled with the redshift.  Error bars of 0.15 mag are shown to
give a sense of the photometric errors that may be expected, excluding
the many possible systematic effects.  \emph{Dashed curve:} The colors
of a blackbody at different redshifts (temperatures) are shown for
reference.  The asterisk marks the color of a 9000K blackbody at
$z=10$, triangles are at intervals of $\Delta z=1$; or equivalently,
at a constant redshift, but with correspondingly different
temperatures (adapted from \cite{MJH06}).
\label{fig:photoz}}
\end{figure}

To the extent that Type II SNe spectra can be represented as a
sequence of blackbodies of different temperatures
(e.g. \cite{DF99}) photometric redshifts will be impossible to
obtain without information about the SN epoch, since temperature and
redshift would be degenerate.  However, local Type IIP SNe show
significant deviations from a blackbody in the UV ($\lambda <
3500$\AA) due to metal-line blanketing in the SN photosphere,
providing spectral signatures that could be used as redshift
indicators, depending on their strength.  In Figure~\ref{fig:photoz},
we show the evolution with redshift of the infrared colors (in bands
accessible with {\it JWST}; see below) of our template SN spectrum,
compared with the color evolution of a blackbody. The figure shows
that the template spectrum deviates significantly from a blackbody. If
the spectrum of the SN is always the same as the template spectrum,
then there are good prospects for obtaining photometric redshifts for
these SNe, at least in the redshift range $z=7-13$.  While the figure
shows that multiple redshifts may be possible at fixed observed
colors, the degenerate solutions would correspond to $z>16$ SNe;
contamination from such high redshift will be mitigated by the fact
that these SNe are likely to be too faint to be detected.  Of course,
more detailed studies of the UV behavior of local SNe, especially
their variety and spectral evolution, will be necessary to confirm the
possible use of photometric redshifts.


\subsubsection{SNe Detectability and Survey Parameters}
\label{sec:survey}

As a specific example for the number of detectable SNe in a future
sample, we consider observations by {\it JWST}.  The relevant
instrument on {\it JWST} is NIRcam,\footnote{See
http://ircamera.as.arizona.edu/nircam for further details.} a
near--infrared imaging detector with a FOV of $2.3'\times4.6'$. A
field can be observed in two filters simultaneously. NIRcam will have
five broadband filters (with resolution $ \lambda / \Delta \lambda
\sim 5$).  We model the filter response as tophat functions with
central wavelengths of 1.5, 2.0, 2.7, 3.5, and 4.5 \micron. For
concreteness, below we will present results only for the 4.5 and 3.5
\micron\ filters, since they are the longest--wavelength {\it JWST}
bands; however, we allow time for imaging in two other bands, if
needed for photometric redshift determinations.  The current estimate
of the {\it JWST} detection threshold at 4.5 \micron~is $\gsim$ 3 nJy
for a 10 $\sigma$ detection and an exposure time of $10^5$ s.  The 3.5
\micron~band is more sensitive, with a detection threshold of $\gsim$
1 nJy for a 10 $\sigma$ detection and an exposure time of $10^5$ s.

\begin{figure}[t]
\centerline{\includegraphics[scale=.45]{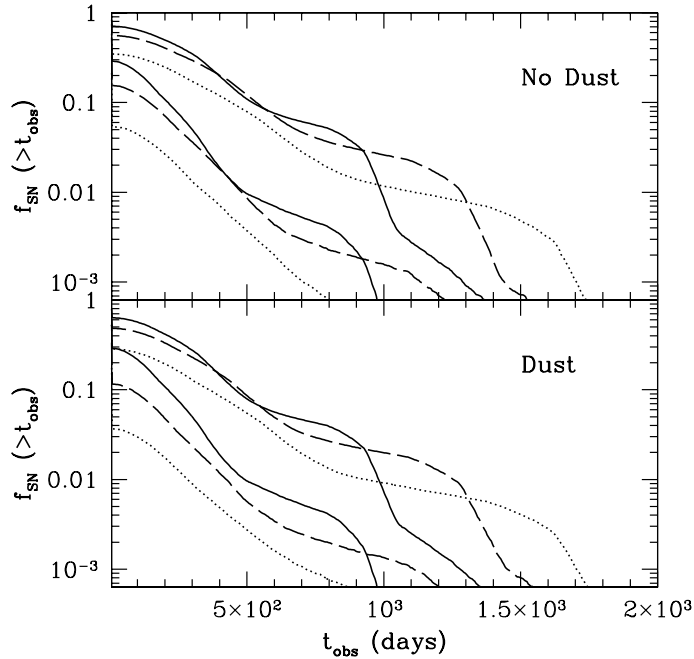}}
\caption{The fraction of SNe which remain visible for an observed
duration of $t_{\rm obs}$ or longer.  The curves correspond to SN
redshifts of $z=7$ ({\it solid curve}), $z=10$ ({\it dashed curve}),
and $z=13$ ({\it dotted curve}).  In each panel, the top set of curves
assumes a flux threshold of 3 nJy (exposure time of $t_{\rm
exp}=10^5$s with the 4.5 \micron~{\it JWST} band), and the bottom set
assumes 9.5 nJy ($t_{\rm exp}=10^4$ s).  The top panel assumes no dust
extinction, while the bottom panel assumes the same dust extinction as
observed within the low--redshift sample \cite{Richardson02} (adapted from
\cite{MJH06}).
\label{fig:f_hist}}
\end{figure}

We show our results for $f_{\rm SN}(>t_{\rm obs})$ in
Figure~\ref{fig:f_hist}.  The solid curves correspond to $z=7$, the
dashed curves correspond to $z=10$, and the dotted curves correspond
to $z=13$.  In each panel, the top set of curves assumes a flux
threshold of 3 nJy (or an exposure time of $t_{\rm exp}=10^5$ seconds
in the 4.5 \micron~band), and the bottom set assumes 9.5 nJy ($t_{\rm
exp}=10^4$ seconds in the 4.5 \micron~band, background dominated).
The top panel further assumes no dust extinction, while in the bottom
panel, we adopt the same dust extinction as in the low redshift sample
\cite{Richardson02}.  Understandably, the distributions get wider as
redshift increases (due to time dilation), but the total visible
fraction, $f_{\rm SN}(>0)$, gets smaller (due to the increase in
luminosity distance).  The double bump feature in some of the curves
corresponds to the plateau of Type IIP SNe lightcurves discussed
above.


\vspace{+0\baselineskip}

As seen in Figure \ref{fig:f_hist}, most of the supernovae that are
bright enough to be visible at all, will remain visible for up to
$\sim$ 1 -- 2 years.  Hence, in order to catch the most SNe, it will
be necessary to have repeat observations of the SN survey fields a few
years apart, to insure that most of the observed SNe will be
identified as new sources or sources that have disappeared.  A time
between observations that is comparable to or larger than the SN
duration will be the optimal strategy for detecting the most SNe
\cite{N03}.  Note that we ignore the time required to collect
reference images, since observations conducted for other programs will
likely provide a sufficient set of such reference images.  However, as
shown in equation~(\ref{eq:SNobs}), we allow time for the field to be
imaged in four different bands, to aid in photometric redshift
determination.

To be more explicit, we find that in order to obtain the largest
number of high--redshift SNe, in general it is a more efficient use of
{\it JWST}'s time to 'tile' multiple fields rather than 'stare' for
extended periods of time ($\gsim 10^4$ s) at the same field
\cite{N03}.  This is because of strong time dilation at these
redshifts.  As can be seen from Figure~\ref{fig:f_hist}, most
detectable SNe will remain above the detection threshold for several
months, even assuming $10^4$ s integration times.  Also evident from
Figure~\ref{fig:f_hist} is that the increase in the total visible
fraction of SNe going from $\texp=10^4$ s to $\texp=10^5$ s, is less
than the factor of 10 increase in exposure time. As a result, a
fiducial 1--yr {\it JWST} survey would therefore detect more SNe using
$\texp=10^4$ s than using $\texp=10^5$ s (see
Figures~\ref{fig:SNobs_45} and \ref{fig:SNobs_35}).  Understandably,
this conclusion does not hold for very high-redshifts, $z\gsim14$,
where SNe are extremely faint, and require very long exposure times to
be detectable.  However, even with such long exposure times, very few
SNe will be detectable at these large redshifts, rendering the use of
longer exposure times unnecessary.

\subsubsection{SNe Detection Rates}
\label{sec:SNe_detect}

\begin{figure}[t]
\centerline{\includegraphics[scale=.5]{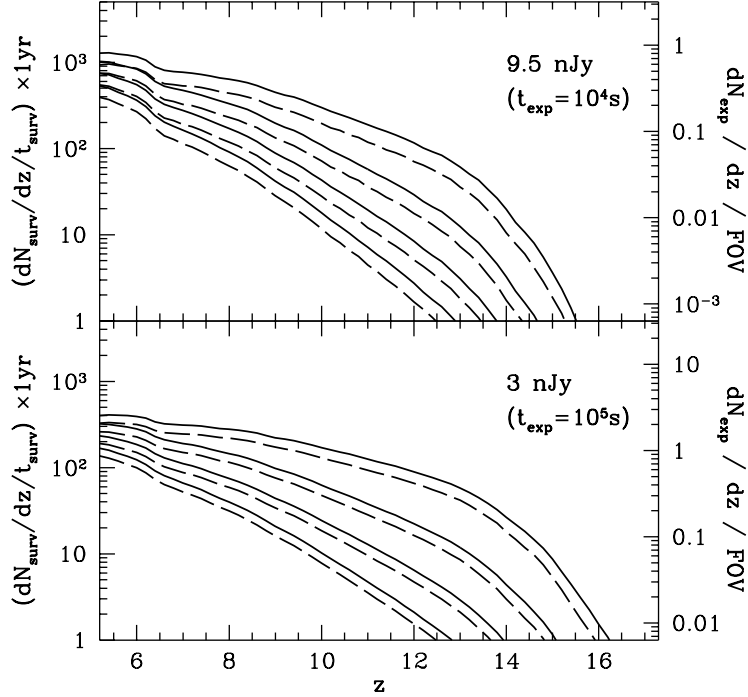}}
\caption{The number of high-redshift SNe detectable with the 4.5
\micron~{\it JWST} filter.  The curves correspond to the same virial
temperature cutoffs as in Figure~\ref{fig:SFR_SNint}.  Solid curves
assume no dust obscuration; dashed curves adopt dust obscuration in
the same amount as observed in the low redshift SNe sample.  The
figure shows results assuming flux density thresholds of 9.5 nJy (or
$\texp = 10^4$ s with {\it JWST}) ({\it top panel}) and 3 nJy (or
$\texp=10^5$ s with {\it JWST}) ({\it bottom panel}).  The right
vertical axis displays the number of SNe per unit redshift per field;
the left vertical axis shows the number of SNe per unit redshift found
in $\tsurv/(2\texp)$ such fields (i.e. the differential version of
eq. (\ref{eq:SNobs}) with $\tsurv=1$ yr).  Reionization should be
marked by a transition from the region bounded by the top two solid
curves to the region enclosed by the bottom three solid curves (or the
analog with the dashed curves if dust is present at the time of
reionization). Adapted from
\cite{MJH06}. \label{fig:SNobs_45}}
\end{figure}

\begin{figure}[t]
\centerline{\includegraphics[scale=.5]{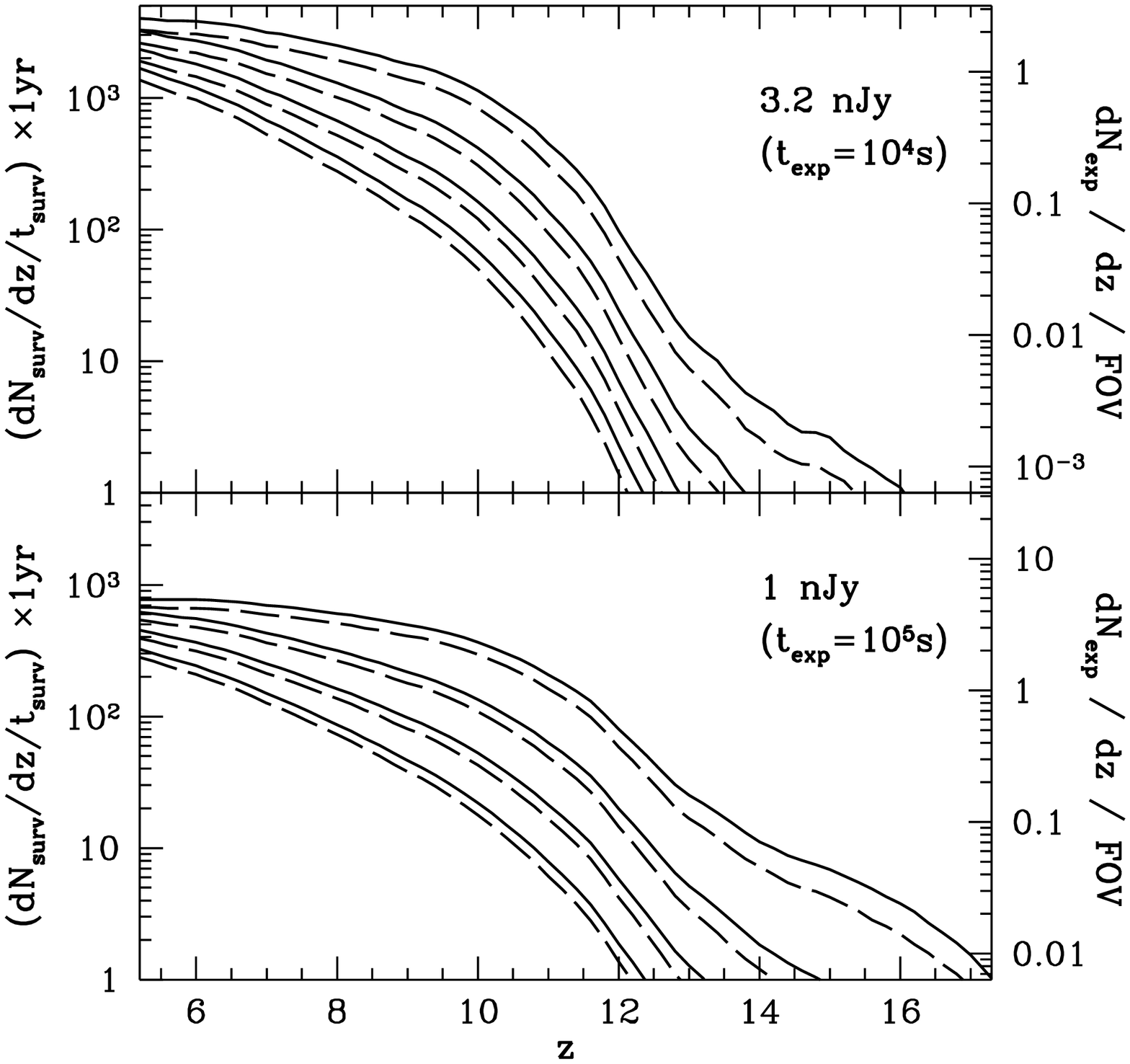}}
\caption{Same as Figure~\ref{fig:SNobs_45}, but with the 3.5
\micron~filter, instead of the 4.5 \micron~one.  Note that we present
results for comparable exposure times, hence the sensitivity
thresholds are three times lower than in Figure~\ref{fig:SNobs_45},
due to the disparate sensitivities of the 3.5 \micron~and 4.5
\micron~filters (adapted from
\cite{MJH06}). \label{fig:SNobs_35}}
\end{figure}

The number of SNe that could be detectable in putative future surveys
are shown in Figures \ref{fig:SNobs_45} and \ref{fig:SNobs_35}.  The
curves correspond to the same virial temperature cutoffs for
star--forming halos as in Figure~\ref{fig:SFR_SNint}.  Solid lines
assume no dust obscuration; dashed lines include a correction for dust
obscuration as discussed above.  Figure \ref{fig:SNobs_45} shows
results assuming flux density thresholds of 9.5 nJy (or $\texp = 10^4$
s with the 4.5 \micron~{\it JWST} filter) ({\it top panel}) and 3 nJy
($\texp=10^5$ s) ({\it bottom panel}).  Figure \ref{fig:SNobs_35}
shows results with the 3.5 \micron~filter assuming equivalent exposure
times: flux density thresholds of 3.2 nJy ($\texp = 10^4$ s with the
3.5 \micron~{\it JWST} filter) ({\it top panel}) and 1 nJy
($\texp=10^5$ s) ({\it bottom panel}).  The right vertical axis
displays the number of SNe per unit redshift per FOV (2.3' $\times$
4.6'); the left vertical axis shows the number of SNe per unit
redshift in a fiducial 1--year survey.  As mentioned above,
reionization should be marked by a transition from the region bounded
by the top two solid curves to the region enclosed by the bottom three
solid curves (or the analog with the dashed curves if dust is present
at the time of reionization).

We note that our expected rates are somewhat higher than those in
\cite{DF99}, a previous study which included SNe lightcurves and
spectra in the analysis.  For example, we find 4 -- 24 SNe per field
at $z\gsim5$ in the 4.5 \micron~filter with $\texp=10^5$ s, compared
to $\sim 0.7$ SNe per field at $z\gsim5$ obtained by \cite{DF99}
(after updating their {\it JWST} specifications to the current
version).  However, they use SFRs extrapolated from the low-redshift
data available at the time, which are not a good fit to recent
high-$z$ SFR estimates \cite{Giavalisco04, Gabasch04, Bunker04}, and
are lower than our $z\gsim5$ SFRs by a factor of 6 -- 40.\footnote{The
possibility that the SFR in a ``Lilly-Madau diagram'' remains flat, or
even increases, at redshifts $z\gsim 5$, owing to star--formation in
early, low--mass halos, is also expected theoretically (see, e.g.,
Fig.1 in \cite{BL02} and associated discussion).}  Taking this factor
into account, their procedure yields 4 -- 27 SNe per field at
$z\gsim5$, which is in excellent agreement with our estimate of 4 --
24 SNe per field at $z\gsim5$.


\subsubsection{Pop--III SNe}
\label{sec:popIII}

\begin{figure}[t]
\centerline{\includegraphics[scale=.45]{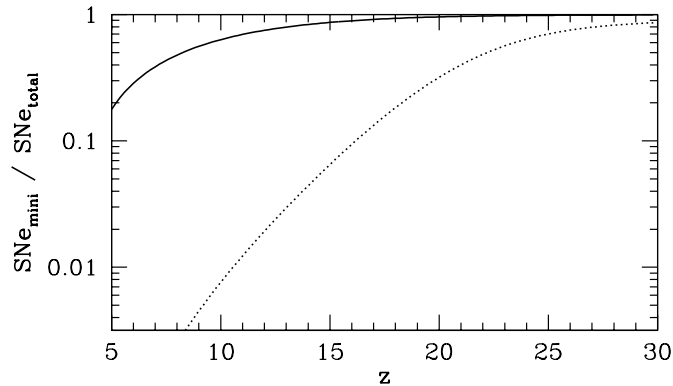}}
\caption{The fraction of SNe whose progenitor stars are located in
minihalos (with virial temperatures of $T_{\rm vir}<10^4$K).  These
SNe may be pair instability SNe, which would be much brighter than the
normal core-collapse SNe used in the computation of SN rates in the
previous figures. The solid curve assumes our fiducial model with a
Salpeter IMF and $\epsilon_{\ast \rm minihalo} = 0.1$; the dotted
curve assumes that each minihalo produces only a single star (and
therefore a single SN) over a dynamical time (adapted from
\cite{MJH06}).
\label{fig:mini_f}}
\end{figure}

As discussed above, high-redshift SNe whose progenitor stars are
formed from metal--free gas within minihalos could be intrinsically
very different from the low-redshift SNe, due to differences in the
progenitor environments (e.g. very low metallicities; \cite{ABN02,
BCL02}).  If such differences could be identified and detected, then
these ``pop--III'' SNe could provide valuable information about
primordial stars and their environments.  Indeed, \cite{WA05}
studied the redshift--distribution of such primordial SNe,
but only briefly addressed the issue of their detectability.  In order to
directly assess the number of such SNe among the hypothetical SNe
samples we obtained here, in Figure~\ref{fig:mini_f}, we plot the
fraction of SNe whose progenitor stars are located in minihalos.  The
solid curve assumes our fiducial model with a Salpeter IMF and
$\epsilon_{\ast \rm minihalo} = 0.1$; the dotted curve assumes that
each minihalo produces only a single star, and hence a single SN, over
a dynamical time (assuming that strong feedback from this star
disrupts any future star formation; as in \cite{WA05}).  The
figure shows that with an unevolving star formation efficiency,
progenitor stars in minihalos would account for $\gsim$ half of the
SNe at $z\gsim9$.  On the other hand, in the extreme case of a single
SNe per minihalo, progenitor stars in minihalos would account for less
than $\sim 1\%$ of the $z\sim 10$ SNe (although at the earliest
epochs, $z\gsim22$, they would still constitute $\gsim$ half of all
SNe).  Given the overall detection rate of several hundred $z\gsim10$
SNe with a 1 year {\it JWST} survey we have found above, even in this
extreme case, Figure~\ref{fig:mini_f} implies that several of these
SNe could be caused by pop--III stars.


\subsection{Detecting Features due to Feedback}

In summary, we find that 4 -- 24 SNe may be detectable from $z\gsim5$
at the sensitivity of 3 nJy (requiring $10^5$ s exposures in a 4.5
\micron~band) in each $\sim 10$ arcmin$^2$ {\it JWST} field.  In a
hypothetical one year survey, we expect to detect up to thousands of
SNe per unit redshift at $z\sim6$.  These rates are high, and, if
reionization produces a fairly sharp features in the reionization
history (with a drop in the SFR by a factor of a few, spread at most
over $\deltazre \sim$ 1 -- 3), we have also shown \cite{MJH06} that
the number of SNe is sufficient to detect the feature out to $z \sim
13$, as well as set constraints on the photo-ionization heating
feedback on low--mass halos at the reionization epoch.  Specifically,
for a wide range of scenarios at $\zre \lsim 13$, the drop in the SNR
due to reionization can be detected at S/N $\gsim$ 3 with only tens of
deep {\it JWST} exposures.  These results therefore suggest that
future searches for high--$z$ SNe could be a valuable new tool,
complementing other techniques, to study the process of reionization,
as well as the feedback mechanism that regulates it.

\section{The Dark Ages as a Probe of The Small Scale Power Spectrum}

Since, as discussed above, GRBs are a rare tracer of the cosmic SFR, a
mere presence of a GRB at, say, $z>10$ will indicate that non-linear
structures exist already at this redshift: the stars that give birth
to the GRBs must form out of gas that collected inside dense dark
matter potential wells.  Structure formation in a cold dark matter
(CDM) dominated universe is ``bottom--up'', with low--mass halos
condensing first.  In the current ``best--fit'' cosmology, with
densities in CDM and ``dark energy'' of $(\Omega_{\rm M},\Omega_{\rm
\Lambda})\approx (0.3,0.7)$ that has emerged from {\it WMAP} and other
recent experiments, DM halos with the masses of globular clusters,
$10^{5-6}\msun$, condense from $\sim 3\sigma$ peaks of the initial
primordial density field as early as $z\sim 25$.  It is natural to
identify these condensations as the sites where the first
astrophysical objects, including the first massive stars, were born.
As a result, one expects to possibly find GRBs out to this limiting
redshift, but not beyond.

With a scale--invariant initial fluctuation power spectrum, the CDM
theory has been remarkably successful, and matched many observed
properties of large--scale structures in the universe, and of the
cosmic microwave background radiation.  However, the power spectrum on
scales corresponding to masses of $M\lsim 10^9~{\rm M_\odot}$ remains
poorly tested.  Some observations have suggested that the standard
model predicts too much power on small scales: it predicts steep cusps
at the centers of dark matter halos, whereas the rotation curves of
dwarf galaxies suggest a flat core; it also predicts more small
satellites than appear to be present in the Local Group (reviewed by,
e.g. \cite{sk01}). Although astrophysical explanations of these
observations are possible, several proposals to solve the problem have
been put forward that involve the properties of dark matter.  These
include self-interacting dark matter \cite{ss00}, adding a repulsive
interaction to gravity \cite{goodman00,peebles00}, the
quantum--mechanical wave properties of ultra--light dark matter
particles \cite{fuzzy}, and a resurrection of warm dark matter
(WDM) models \cite{bode01}.

By design, a common feature of models that attempt to solve the
apparent small-scale problems of CDM is the reduction of fluctuation
power on small scales. The loss of small-scale power modifies
structure formation most severely at the highest redshifts, where the
number of self--gravitating objects is reduced. {\em In each model,
there exists a redshift beyond which the number of GRBs (or any other
object) is exponentially suppressed; a detection of a GRB beyond this
redshift can be used to rule out such models.} As an example,
ref.\cite{bho01} showed that in the case of WDM models, invoking a WDM
particle mass of $\sim 1$keV (the mass needed to solve the problems
mentioned above), it becomes difficult to account for the reionization
of the universe by redshift $z\sim 6$, due to the paucity of ionizing
sources beyond $z=6$.

GRBs, if discovered in sufficient numbers at $z>6$, have the potential
to improve this constraints significantly, since increasingly higher
redshifts probe increasingly smaller scales.  The constraints, in
particular, that the detection of distant GRBs would place on
structure formation models with reduced small--scale power was
quantified in \cite{MPH05}.  In this work, we computed the number of
GRBs that could be detectable by the {\it Swift} satellite at high
redshifts ($z\gsim 6$), assuming that the GRBs trace the cosmic star
formation history, which itself traces the formation of non--linear
structures. Simple models of the intrinsic luminosity function of the
bursts were calibrated to the number and flux distribution of GRBs
observed by the {\it Burst And Transient Source Experiment
(BATSE)}. Under these assumptions, the discovery of high--$z$ GRBs
would imply strong constraints on models with reduced small--scale
power. For example, a single GRB at $z\gsim 10$, or, alternatively, 10
GRBs at $z\gsim 5$, discovered within a two--year period, would rule
out an exponential suppression of the power spectrum on scales below
$R_c=0.09$ Mpc (exemplified by warm dark matter models with a particle
mass of $m_x=2$ keV).  Models with a less sharp suppression of
small--scale power, such as those with a red tilt or a running scalar
index, $n_s$, are more difficult to constrain, because they are more
degenerate with an increase in the power spectrum normalization,
$\sigma_8$, and with models in which star--formation is allowed in
low--mass minihalos.  We find that a tilt of $\delta n_s\approx 0.1$
is difficult to detect; however, an observed rate of 1 GRB/yr at $z
\gsim 12$ would yield an upper limit on the running of the spectral
index, $\alpha\equiv dn_s/d\ln k > -0.05$.

\begin{acknowledgement}
This article draws on joint work with numerous colleagues over the
past several years. I would like to thank all my collaborators, but
especially Andrei Mesinger, Ben Johnson, Greg Bryan, Mark Dijkstra,
and Roban Kramer, whose recent works were particularly emphasized
here.  The work described in this review was supported by NASA, the
NSF, and by the the Pol\'anyi Program of the Hungarian National Office
for Research and Technology (NKTH).
\end{acknowledgement}
%

%
%
%

\end{document}